\documentclass[aps,prx,floatfix,superscriptaddress,showpacs,twocolumn,nofootinbib,longbibliography]{revtex4-1}

\usepackage{graphicx}
\usepackage{float}
\usepackage{amsmath}
\usepackage{amssymb}
\usepackage{amsfonts}
\usepackage{amsthm}
\usepackage{mathtools}
\usepackage[unicode=true, breaklinks=false, pdfborder={0 0 1}, backref=false, colorlinks=true, linkcolor=blue, citecolor=blue, urlcolor=blue]{hyperref}
\usepackage{color}
\usepackage{soul}
\usepackage{textcomp}
\usepackage{bm}
\usepackage{dsfont}
\usepackage{relsize}
\usepackage{braket}
\usepackage{enumitem}   
\usepackage{mathrsfs}
\usepackage{cleveref}
\usepackage{bigints}
\usepackage{framed}
\usepackage[makeroom]{cancel}
\usepackage{comment}

\usepackage{soul}

\theoremstyle{plain}

\numberwithin{obs}{section}

\definecolor{Blue}{rgb}{0,0,1}
\definecolor{Red}{rgb}{1,0,0}
\definecolor{Green}{rgb}{0,1,0}
\definecolor{darkgreen}{rgb}{0,.7,0}
\definecolor{Purp}{rgb}{.2,0,.2}
\definecolor{white}{rgb}{1,1,1}

\newcommand{\comments}[1]{}
\newcommand{\ba}{\begin{align}}
\newcommand{\ea}{\end{align}}

\newcommand{\ee}{{\mathcal{E}}}

\newcommand{\Tr}[1]{\text{Tr}\left[#1\right]}

\newcommand{\rom}[1]{\uppercase\expandafter{\romannumeral #1\relax}}

 \def\be{\begin{eqnarray}}
\def\ee{\end{eqnarray}}

\begin{document}

\title{
Fundamental limits in Bayesian 
thermometry and  attainability via adaptive strategies
}
\author{Mohammad Mehboudi}
    \email{mohammad.mehboudi@unige.ch}
    \affiliation{D\'{e}partement de Physique Appliqu\'{e}e, Universit\'{e} de Gen\`{e}ve, 1211 Geneva, Switzerland}
\author{Mathias R. J{\o}rgensen}
    \email{matrj@fysik.dtu.dk}
    \affiliation{Department of Physics, Technical University of Denmark, 2800 Kongens Lyngby, Denmark}
\author{Stella Seah}
    \affiliation{D\'{e}partement de Physique Appliqu\'{e}e, Universit\'{e} de Gen\`{e}ve, 1211 Geneva, Switzerland}    
\author{Jonatan B. Brask}
    \affiliation{Department of Physics, Technical University of Denmark, 2800 Kongens Lyngby, Denmark}
\author{Jan Ko\l{}ody\'{n}ski}
    \affiliation{Centre for Quantum Optical Technologies, Centre of New Technologies, University of Warsaw, 02-097 Warsaw, Poland}
\author{Mart\'{i} Perarnau-Llobet}
    \email{marti.perarnaullobet@unige.ch}
    \affiliation{D\'{e}partement de Physique Appliqu\'{e}e, Universit\'{e} de Gen\`{e}ve, 1211 Geneva, Switzerland}    
\begin{abstract}
    We investigate the limits of thermometry using quantum probes at thermal equilibrium within the Bayesian  
    approach. We consider the possibility of engineering interactions between the probes in order to enhance their sensitivity, as well as feedback during the measurement process, i.e., adaptive protocols. 
    On the one hand, we obtain an ultimate bound on thermometry precision in the Bayesian setting, valid for arbitrary interactions and measurement schemes, which lower bounds the error with a quadratic (Heisenberg-like) scaling with the number of probes. 
    We develop a simple adaptive strategy that can saturate this limit. On the other hand, 
    we  derive 
    a no-go theorem for non-adaptive protocols that does not allow for better than linear (shot-noise-like) scaling  even if one has unlimited control over the probes, 
    namely access to arbitrary many-body interactions. 
\end{abstract}
\maketitle
%
%
{\it Introduction.---}Preparing quantum systems at low temperatures is an essential task for development of quantum technologies 
\cite{Celi_2016,RevModPhys.80.885,bloch2012quantum}.
Measuring temperature precisely is necessary to validate cooling  and ensure the performance of quantum protocols, and has been demonstrated in cutting-edge experiments
\cite{leanhardt2003cooling,PhysRevX.10.011018,PhysRevX.10.041054,olf2015thermometry,PhysRevLett.96.130404,ronzani2018tunable,PhysRevLett.103.223203,tan2017quantum,adam2021coherent}; it is however challenging. Due to the scarcity of thermal fluctuations at such low temperatures, 
the relative error on thermometry 
can be enormous. Moreover, the fragility of quantum systems requires additional forward planning to minimise disturbance while maximising the information obtained. The theory of quantum thermometry is built to address these pivotal challenges~\cite{Mehboudi_2019,de2018quantum}.

Quantum thermometry finds fundamental limits on precision \cite{correa2015individual,Paris_2015,Potts2019fundamentallimits,PhysRevResearch.2.033394} and designs protocols to achieve them in different platforms~\cite{PhysRevLett.122.030403,PhysRevA.95.022121,PhysRevLett.103.170404,PhysRevLett.125.080402}, and improve them thanks to quantum correlations \cite{PhysRevLett.123.180602,Campbell_2017}, coherence \cite{PhysRevA.91.012331,razavian2019quantum}, many-body interactions and criticality \cite{de2016local,PhysRevB.98.045101,Mehboudi_2015,PhysRevA.103.023317,Latune_2020,planella2020bath} or other resources \cite{PhysRevA.96.062103,PhysRevLett.119.090603}. 
To date, such enhancements have been developed in the context of 
local thermometry, aiming at designing a thermometer that detects the smallest temperature variations around a known temperature~\cite{Mehboudi_2019,de2018quantum}. In many practical situations, however, one might not know the temperature accurately beforehand. Rather, one has limited prior knowledge about the temperature of the sample.
Under such circumstances, 
Bayesian 
estimation is a more suitable approach,
and has been the subject of a few recent studies~\cite{rubio2020global,alves2021bayesian,PhysRevA.104.052214}. 

The goal of this work is to set the ultimate bounds of
Bayesian equilibrium thermometry, and to develop adaptive  strategies to saturate them. It is insightful to first recall analogous results in the local approach to 
equilibrium thermometry~\cite{Mehboudi_2019,de2018quantum}. 
Within such a framework---contrary to dynamical approaches where the probe evolves according to some predefined model parametrised by the temperature~\cite{Kiilerich2018,Seah2020}, e.g.~a superconducting qubit in radiometry~\cite{Wang2021}---the probe always thermalises to the temperature of the sample whose value is known \emph{a priori}. 
In that case, for any unbiased 
estimator ${\tilde \theta}$  of the temperature $\theta_0$, the mean square error is inversely proportional to the heat capacity of the probe: $\Delta {\tilde \theta} \propto 1/C$~\cite{Old_PRE,Paris_2015,correa2015individual,Mehboudi_2015}. For $n$-body probes, $C$~can scale super-extensively with~$n$ 
in the vicinity of a critical point, with the ultimate bound~$C \approx n^2/4$~\cite{correa2015individual,Marcin2018}---a quadratic scaling with the number of resources reminiscent of the Heisenberg scaling in quantum metrology~\cite{giovannetti2004quantum}. Here, we  show that similar bounds hold in the Bayesian approach, but adaptive strategies are needed to saturate them, contrary to the local case. In fact, we  prove that any non-adaptive strategy necessarily leads to
$\Delta {\tilde \theta} \propto 1/n$  for sufficiently large~$n$---i.e., a 
shot-noise-like scaling~\cite{giovannetti2004quantum}---a no-go result that holds even when arbitrary control over the $n$-body probe Hamiltonian is allowed. Thus, adaptive measurement strategies are a crucial ingredient for optimal thermometry whenever the temperature value is \textit{a priori} not perfectly known.  

\begin{figure}
    \centering
    \includegraphics[width=.85\columnwidth]{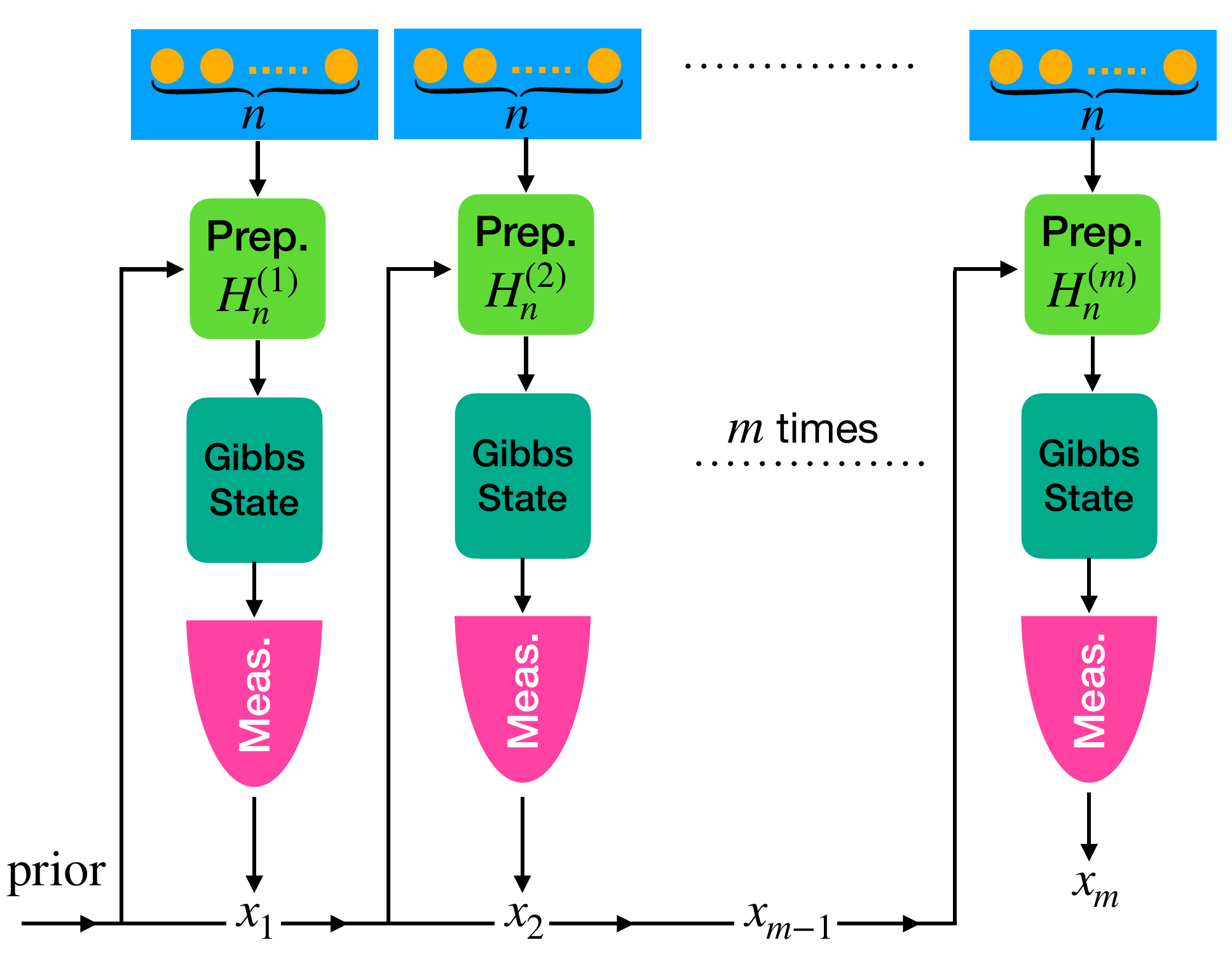}
    \caption{Schematic representation of the adaptive scenario. A total of $N$ probes are used in groups of $n$ to estimate the temperature of the sample, $\theta_0$. Initially, our prior temperature distribution is given by $p(\theta)$, according to which we choose the Hamiltonian of the first $n$ probes to be $H_n^{(1)}$ that minimises the expected mean square logarithmic error.
    The probes interact and thermalise with the sample followed by an energy measurement, yielding an outcome, say $x_1$. Our knowledge about the temperature will be reflected in the posterior distribution $p(\theta|x_1)$. This will be used as the prior for the second round---in order to find the optimal Hamiltonian ~$H_n^{(2)}$. This process is repeated~$m=N/n$ times. In contrast, in the non-adaptive scenario the Hamiltonian is fixed~$H_n^{(k)}=H_n~\forall k$. }
    \label{fig:set-up}
\end{figure}

{\it Preliminaries and setup.---} 
%
We consider estimation of the temperature $\theta_{0}$ of a (possibly macroscopic) sample given some  prior distribution $p(\theta)$ reflecting our initial knowledge on~$\theta_{0}$.
We assume we have at our disposal $N$ copies of a $d$-dimensional system that we use as probes, which are much smaller than the sample. When put in contact with the sample, we assume that the probes eventually  reach thermal equilibrium at temperature~$\theta_{0}$.  By measuring them we infer $\theta_{0}$. This corresponds to the framework of equilibrium thermometry, which is by nature robust~\cite{Mehboudi_2019,de2018quantum}. 
In order to establish fundamental bounds, we  assume full control on the Hamiltonian of the probes, and in particular the ability to make them interact. Therefore, alternatively one can  think of a $d^N$-dimensional probe, which constitutes our resource. 


The thermometry process is divided into $m$ rounds, each involving $n=N/m$ probes.  Every round consists of: (I) preparation of the $n$-body probe, (II) interaction with the sample and thermalisation, (III) measurement/data acquisition, and (IV) data analysis (see Fig.~\ref{fig:set-up}).  In the first round,  
we start by engineering  the Hamiltonian~$H_n^{(1)}$ of the  $n$-body probe into any desired configuration based on the prior distribution~$p(\theta)$. That is, we arrange the energy distribution of the $n$-body probe to become most sensitive to the relevant temperature range.   
Next, in step~(II), this   $n$-body system is put in contact with the sample, and reaches thermal equilibrium with it. Therefore, it can be described by the Gibbs state $\omega_{\theta_0}(H_n^{(1)}) \coloneqq \exp[-H_n^{(1)}/{\theta}_{0}]/Z$, with $Z={\rm Tr} (\exp[-H_n^{(1)}/{\theta}_{0}])$  the partition function.
Then, in step (III),  a measurement is performed that yields an outcome $x_1$. We focus on energy measurements since they are optimal as the Gibbs state is diagonal in the energy basis. 
In the data analysis  (step (IV)), the posterior distribution is obtained through Bayes' rule:
\begin{equation}
    p(\theta | x_1) = \frac{p(x_1 | \theta) p(\theta)}{p(x_1)} ,
\end{equation}
where $p(x| \theta)$ is the likelihood function (which depends on the temperature \textit{and} the Hamiltonian), $p(\theta)$ is the prior distribution on $\theta$,
and 
$p(x)=\int\!{\rm d}\theta\,p(\theta)\, p(x|\theta)$
is the outcome probability. The next round proceeds in an analogous way, but replacing the prior~$p(\theta)$ by~$p(\theta | x_1)$ and $H_n^{(1)}$ by~$H_n^{(2)}$. Likewise, in round $k > 1$,   $p(\theta)$ is replaced by $p(\theta | {\bf x}_{k-1})$ with ${\bf x}_{k-1}\equiv \{ x_{k-1},...,x_2,x_1 \}$ and $H_n^{(1)}$ is replaced by $H_n^{(k)}$. 
Such a strategy is adaptive since $H_n^{(k)}$ depends on ${\bf x}_{k-1}$. In contrast, a non-adaptive strategy satisfies $H_n^{(k)}=H_n$ $\forall k$, where $H_n$ is chosen according to the initial prior $p(\theta)$ only.
At the end of the thermometry process (round $m$), the final estimate ${\tilde \theta}({\bf x}_m)$ of $\theta_{0}$ is computed.


%

In order to gauge the quality of the estimator, we need to introduce 
an error quantifier that describes how far ${\tilde \theta}$ is from $\theta_{0}$, on average. 
A natural measure which is suitable for equilibrium probes is the expected mean square logarithmic error ($\text{\footnotesize{EMSLE}}$) (see~\cite{rubio2020global} for justification and the accompanying paper~\cite{Mathias_accompany} for a deeper analysis and generalisation)
\begin{align}
    \hspace{-1.5mm}
    \text{\footnotesize{EMSLE}}\coloneqq \hspace{-1mm} \int   {\hspace{-1mm} \rm d}\theta \hspace{0.5mm} p(\theta) \hspace{-1mm} \int{\hspace{-1mm}  \rm d}{\bf x}_m \hspace{0.5mm} p({\bf x}_m|\theta) 
    \ln^2 \left[\frac{{\tilde \theta}({\bf x}_m)}{\theta} \right],
    \label{mle}
\end{align}
with $d{\bf x}_m \coloneqq dx_m...dx_1$. 
Moreover, 
\begin{align}
    {\tilde \theta}({\bf x}_m) = \exp\left[\int {\rm d} \theta \frac{p(\theta) p({\bf x}_m|\theta)}{p({\bf x}_m)} \ln \theta \right],
\end{align}
is the optimal temperature estimator, i.e., it minimises $\text{\footnotesize{EMSLE}}$~\cite{rubio2020global}. 

We wish to find lower bounds for $\text{\footnotesize{EMSLE}}$, as well as optimal strategies to saturate them, for both  adaptive and non-adaptive measurements.  More precisely, our aim is to minimise $\text{\footnotesize{EMSLE}}$ as a function of the number $N$ of probes, with $N=m n$. 
We will pay particular attention to the relevant case where $m\gg 1$ is large (asymptotic regime) but $n$ is limited due to e.g.~experimental limitations on the amount of probes that can be collectively processed.  In this case, we will focus on the scaling of $\text{\footnotesize{EMSLE}}$ with $n$ for a fixed but large~$m$.

{\it Main results.---}Our main results are (i)~an ultimate precision limit for Bayesian thermometry that holds for both adaptive and non-adaptive strategies, which in principle allows for a 
quadratic (Heisenberg-like) scaling with~$n$, (ii)~a no-go theorem that forbids super-extensive scaling in any non-adaptive scenario, and (iii)~an adaptive strategy that reaches the ultimate limit. 
These results are derived in what follows (some technical details are given in the Appendix). 

Given the prior $p(\theta)$, and by utilising the Van Trees inequality~\cite{van2004detection,trees2007bayesian} we construct a lower bound on the estimation error after $m$ rounds
\begin{align}\label{eq:bcrb_adpt_main}
    &\text{\footnotesize{EMSLE}}^{-1}
    \leqslant Q[p(\theta)] \nonumber\\
    &+\sum_{k=1}^{m} \int d\textbf{x}_{k-1} \hspace{0.5mm} p(\textbf{x}_{k-1})  \int d\theta 
   \hspace{0.5mm} \hspace{0.5mm} p(\theta|{\bf x}_{k-1})
  \hspace{0.5mm} C(\theta;H_n^{(k)}),
\end{align}
where $p(\theta|{\textbf{x}_{0}}) = p(\theta)$, $p({\textbf{x}_{0}})= p({\textbf{x}_{0}}|\theta)=1$, and $\int d{\textbf{x}_{0}} =1$ are introduced to compress our notation. Here, $Q[p(\theta)]$ quantifies the prior information and reads
\begin{equation}
    Q[p(\theta)] \coloneqq \int d\theta \ p(\theta) \left[1+ \theta \partial_{\theta} \log p(\theta) \right]^{2}.
\end{equation}
The second term quantifies the information acquired through all measurements. It also establishes a connection to the quantum Fisher information through  its proportionality to the heat capacity~\cite{correa2015individual}.
The heat capacity of the probe at round $k$ of the measurement is denoted
$C(\theta;H_n^{(k)})$, with the Hamiltonian $H_n^{(k)}$ designed according to the prior \textit{and} the information acquired so far. Recall that, by definition, $C(\theta;H_n)\coloneqq\partial_{\theta} E(\theta;H_n)$ where $E(\theta;H)={\rm Tr}[H \omega_\theta (H)]$ is the energy of the probe at thermal equilibrium. 
To bound Eq.~\eqref{eq:bcrb_adpt_main}, we first 
define the maximum of the integrand over $\{ H_n^{(k)}\}_{k=1}^{m}$ for a specific trajectory ${\bf x}_m$: 
\begin{align}\label{eq:Gamma_main}
    \Gamma({\bf x}_{m}) \coloneqq & \max_{{\{ H_n^{(k)}\}}_{k}}\,\sum_{k=1}^{m} \int d\theta 
   \hspace{0.5mm} p(\theta|{\bf x}_{k-1})C(\theta;H_n^{(k)})  \nonumber\\
   &\leqslant\sum_{k=1}^{m} \int d\theta 
   \hspace{0.5mm} p(\theta|{\bf x}_{k-1}) C_D 
   =m C_D
\end{align}
where $C_D\coloneqq \max_{H_n}C(\theta;H_n)$, i.e., the maximum heat capacity of an $n$-body probe. In the last line we used that $C_D$ is independent of $\theta$  (see  \cite{correa2015individual} and the Appendix for the explicit expression of $C_D$). 
Furthermore, we have $C_D \approx \frac{n^2}{4}\log^2 d$, for large enough $n$. 
Putting everything together, we obtain from \eqref{eq:bcrb_adpt_main}: 
\begin{align}\label{eq:ultimate_main}
   \text{\footnotesize{EMSLE}}^{-1}&\leqslant Q[p(\theta)] + m C_D
    \nonumber\\ 
   & \overset{n\gg 1}{\approx}Q[p(\theta)] + m\frac{n^2}{4}\log^2 d.
\end{align}
This gives an ultimate bound on Bayesian thermometry [Result (i)], which both adaptive and non-adaptive strategies should respect. 
This bound implies that any Bayesian thermometry protocol is ultimately limited by  a quadratic  Heisenberg-like scaling. 

The ultimate bound \eqref{eq:ultimate_main} becomes tight and can be saturated by adaptive strategies 
in the regime $m\gg 1$ (see results below). However,  non-adaptive strategies fail to saturate it, and in fact 
$\text{\footnotesize{EMSLE}}^{-1}$
can increase at most linearly with $n$ [Result (ii)]
\begin{align}\label{eq:non_adapt_nogo}
 \hspace{-1mm} \text{\footnotesize{EMSLE}}^{-1}
        \overset{\mathrm{non-adaptive}}{\leqslant} Q[p(\theta)] + f[p(\theta)] mn\log\!d, 
\end{align}
where $f[p(\theta)] = \int_{\cal R} d \theta \ [-\partial_{\theta}p(\theta)] \theta$ is a functional of only the prior distribution,   and ${\cal R}$ is the temperature domain where $\partial_{\theta}p(\theta)\leqslant 0$.
This result is rigorously proven in the Appendix, but let us provide some intuition.
%
It is already noted in the literature that engineered probes for thermometry show enhanced sensitivity only in a small temperature range~$\Delta$~\cite{correa2015individual,Mehboudi_2019,Campbell2018,RomnAncheyta2019,mok2020optimal}. Finite-size scaling theory hints that if $C \propto n^{1+\alpha}$, then $\Delta \propto n^{-\gamma}$ with $\gamma \geqslant \alpha$ in order to ensure  
that the energy density of an equilibrium state remains finite~\cite{huang2009introduction}.  This implies that, for any  $p(\theta)$ with a finite width (independent of $n$),  the term $\int {\rm d}\theta \hspace{0.5mm} p(\theta) C(\theta)$ in Eq.~\eqref{eq:bcrb_adpt_main} grows  at most linearly with $n$ for sufficiently large $n$. In other words, optimal $n$-body probes require  priors  with a   width smaller than $\mathcal{O}(1/n)$ to obtain super-linear scaling, and conversely a finite  width in $p(\theta)$  will eventually kill any super-linear scaling.  The  no-go result \eqref{eq:non_adapt_nogo} makes this intuition rigorous. 

The above reasoning  also explains why adaptive protocols can potentially saturate \eqref{eq:ultimate_main}. By  updating the prior $p(\theta)$ to the posterior $p(\theta | {\bf x}_{k-1})$ in each step of the process ($k=1,...,m$), it can stay inside the optimal region for sufficiently large $m$, thus enabling super-linear precision. 
This also suggests using  optimal probes for local thermometry as an \textit{ansatz} for the Bayesian thermometry with adaptive strategies. The optimal thermometer in the local scenario is an effective two-level system with $d^n-1$-fold degeneracy in the excited state~\cite{correa2015individual}.
Although this Hamiltonian is useful to obtain fundamental bounds~\cite{correa2015individual} it involves $n$-body interactions and is hence highly complex for $n\gg 1$. Nonetheless,  it can be well approximated through two-body interactions by the method developed in~\cite{chancellor2016direct} and, furthermore, it can be effectively realised   with a few-fermionic mixture confined in a one-dimensional harmonic trap~\cite{Marcin2018}.
Motivated by this progress, at the $k$th round we restrict to the class of Hamiltonians $H_n^{(k)}$ with the aforementioned two-level structure, and tune the energy gap to minimise the
\text{\footnotesize{EMSLE}}~\eqref{mle}. 
As we show in the example below,  we can achieve a quadratic scaling with $n$ and saturate~\eqref{eq:ultimate_main} using this strategy [Result~(iii)].
\begin{figure*}
	\begin{tabular}{c}
\includegraphics[width=\columnwidth]{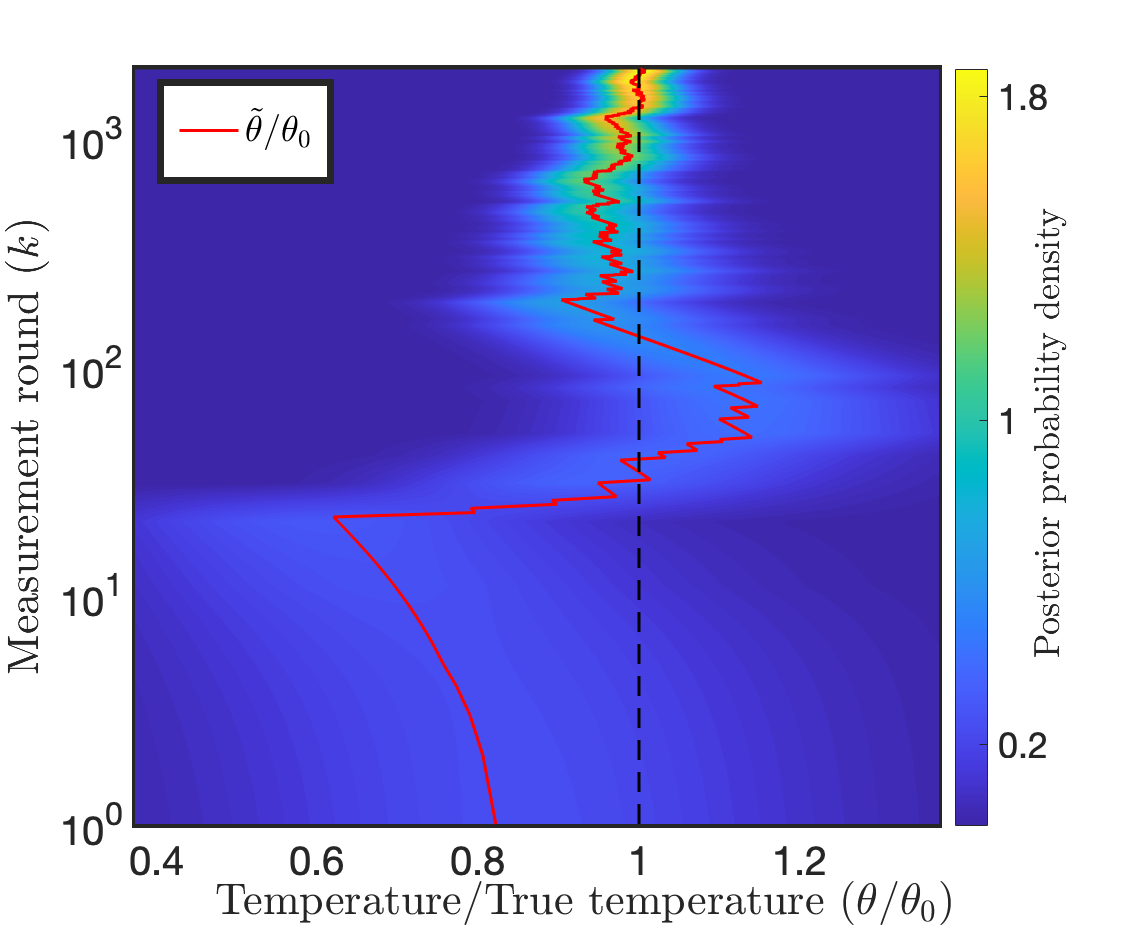}
\includegraphics[width=\columnwidth]{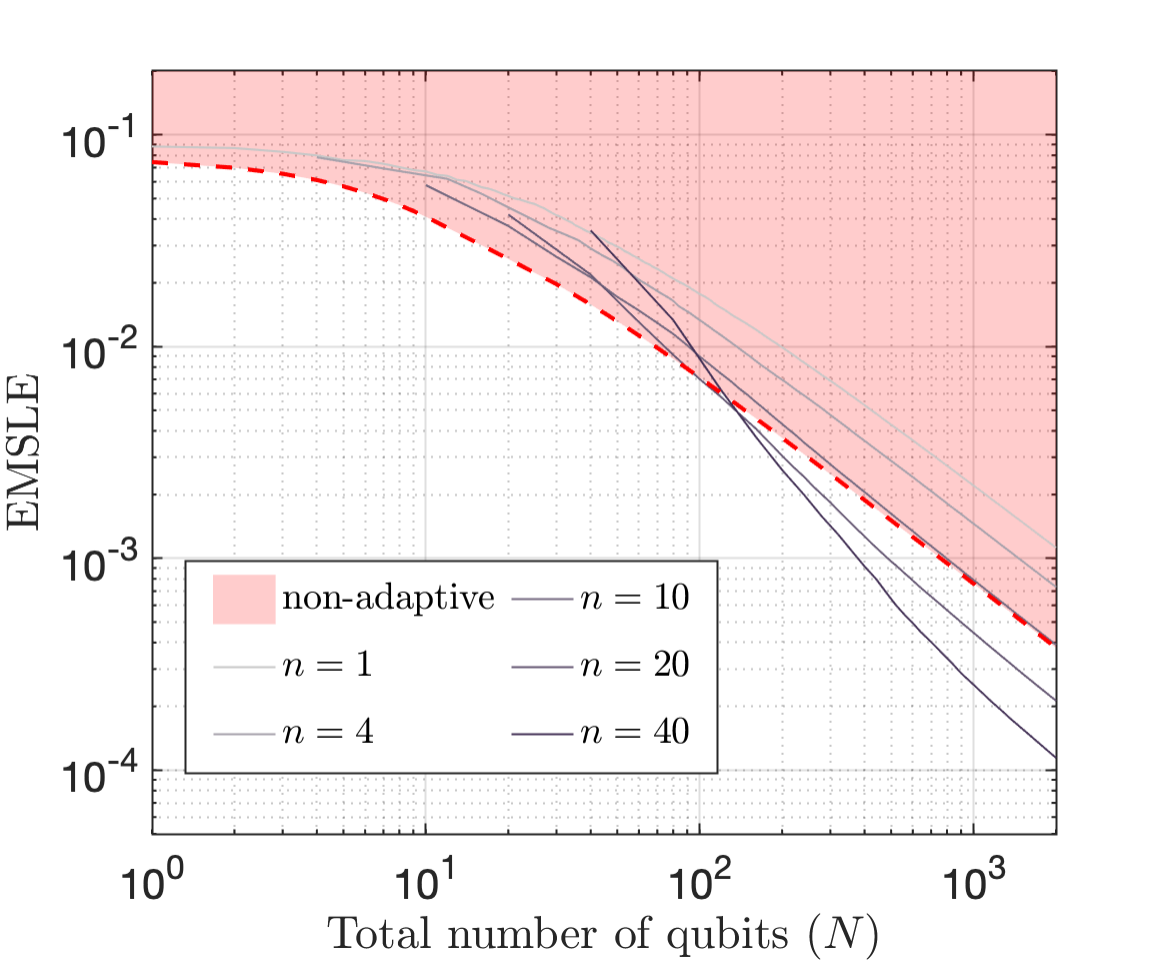}
	\end{tabular}
    \caption{Left---Contour plot of the prior versus the measurement round $k\in \{1,\dots, m\}$ (logarithmic scale), and temperature normalised to its true value $\theta/\theta_{0}$. The red trajectory shows the ratio between the estimated temperature and the true temperature ${\tilde \theta}/\theta_0$. As $k$ increases, the prior sharpens around the true temperature, and ${\tilde \theta}/\theta_{0}$ approaches one. Here, we have set $n=1$, $\alpha=1$, $\theta_{\min}=1 $, and $\theta_{\max} = 10$ in arbitrary units.
    Right---Loglog plot of 
    the expected mean square logarithmic error (\text{\footnotesize{EMSLE}}) attained by the
    adaptive strategy vs.\ the total number of qubits $N$. Dark solid lines represent different values of $n$. They show that, for sufficiently large $N$, the bigger $n$ is the smaller the error can get. The red-dashed line is the (not necessarily tight) bound on non-adaptive strategies: only the shaded area can be achieved using non-adaptive protocols. One can cross the border with adaptive strategies for $n>10$.}
    \label{fig:num_2000_error}
\end{figure*}
\begin{figure}
    \includegraphics[width=\columnwidth]{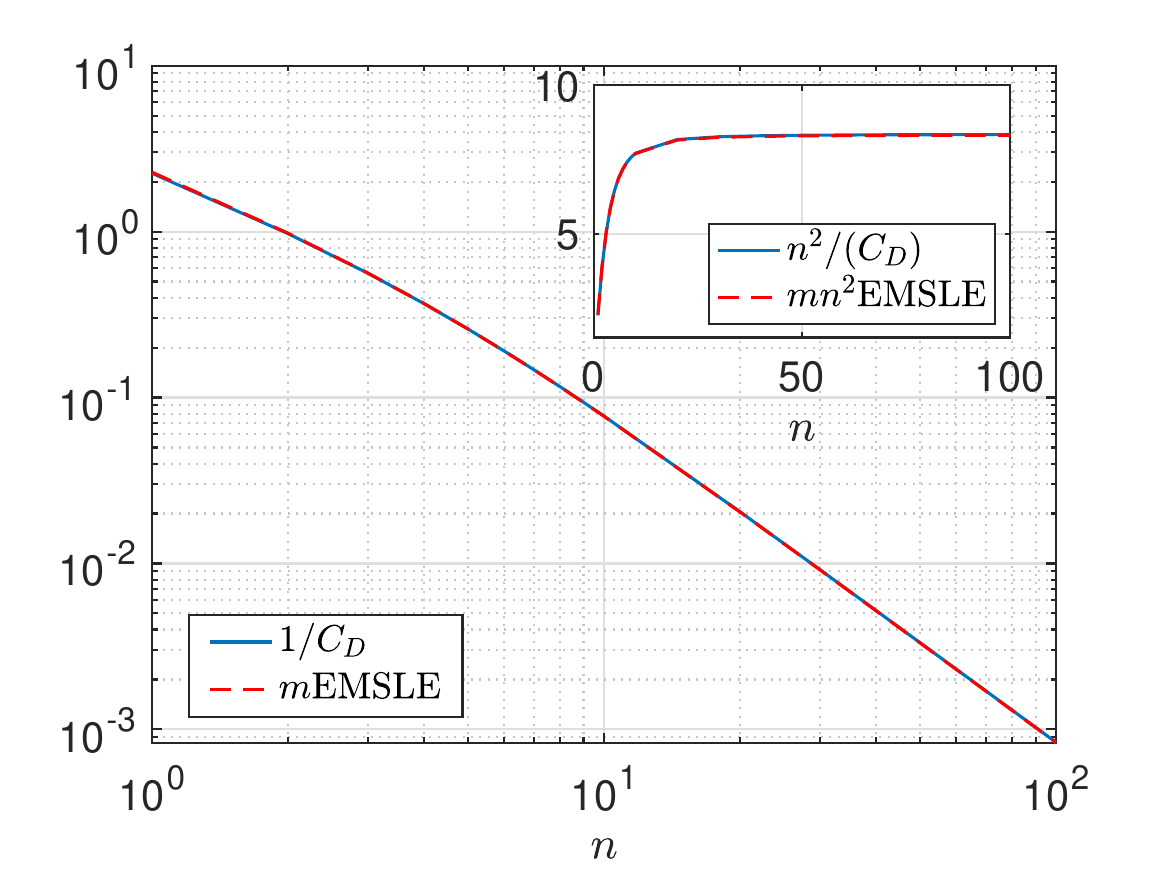}
    \caption{(Dashed red)
    Loglog plot of the normalised 
    expected mean square logarithmic error (\text{\footnotesize{EMSLE}})
    after $m$ rounds of the adaptive scheme---for sufficiently large $m$, here $m=2\times 10^3$---vs $n$. This shows that for large enough $n$ the error vanishes quadratically with $n$, which can be better seen from the inset. 
    (Blue)
    The minimum achievable \text{\footnotesize{EMSLE}} given by the r.h.s of \eqref{eq:ultimate_main}. The perfect agreement shows the efficiency of the proposed adaptive protocol.}
    \label{fig:error_vs_groupedin}
\end{figure}

{\it Case study.---}The results presented here are valid for a broad class of priors, but in what follows we stick to a specific choice in order to illustrate their usage. 
In any relevant application of thermometry, 
the temperature is known a priori to lie within a certain range, i.e., $\theta_{\min}\leqslant \theta_{0} \leqslant \theta_{\max}$. We use a family of probability distributions that are suitable in this case and were proposed in~\cite{Yan2018}:
\begin{equation}
\label{eq:prior}
    p(\theta) = \frac{1}{k_{\alpha} (\theta_{\max}-\theta_{\min})} \left[ e^{\alpha \sin^{2}\left( \pi \frac{\theta-\theta_{\min}}{\theta_{\max}-\theta_{\min}} \right) } - 1 \right]
\end{equation}
with
\begin{equation}
    k_{\alpha} \coloneqq e^{\alpha/2} I_{0}(\alpha/2) - 1 ,
\end{equation}
where $I_{0}$ is the modified Bessel function of the first kind.
In the limit $\alpha\rightarrow -\infty$ the above prior becomes a constant, while in the limit $\alpha\rightarrow 0$ we have $p(\theta)\propto \sin^{2}(2\theta)$. 


The adaptive strategy works as follows. We consider as a resource $N$ qubits, which are divided in $m$ groups of  $n$ qubits. In each group, the $n$-qubit Hamiltonian is engineered to become a two-level system with degeneracy 
$(2^{n}-1)$ and with  a tunable gap $\epsilon$ 
In the first round, we tune the gap to $\epsilon^{(1)}$ to minimise the single shot $\text{\footnotesize{EMSLE}}$, that is we set $m=1$ in \eqref{mle}.
Then, we measure the energy of the system. Given the outcome $x_1$ is observed, we update the prior to $p(\theta) \rightarrow p(\theta | x_1)$, and implement the same procedure to choose~$\epsilon^{(2)}$ in the second round (i.e., we minimise~\eqref{mle} replacing $p(\theta) \rightarrow p(\theta | x_1)$). This process is repeated until all probes are used.

In our simulations, we apply the adaptive process for a given  $\theta_{0}$ sampled from $p(\theta)$, which yields a trajectory as illustrated in the left panel of Fig.~\ref{fig:num_2000_error}. We see
that the prior peaks around the true temperature as   $k$ increases, and the estimated temperature gets closer to the true temperature, i.e., ${\tilde \theta}/\theta_{0}\to 1$. The average over a large amount of trajectories enables us to  compute $\text{\footnotesize{EMSLE}}$ in Eq.~\eqref{mle} with high accuracy (in the numerical simulations, we consider $\mathcal{O}(1000/m)$ trajectories, which ensures convergence).  
In the right panel of Fig.~\ref{fig:error_vs_groupedin} we plot $\text{\footnotesize{EMSLE}}$ in the  adaptive scenario 
for various values of $n$, benchmarked against  the no-go bound for non-adaptive scenarios---only the shaded area can be accessed by non-adaptive strategies given any~$n\leqslant N$. 
We see that as $n$  increases the error gets smaller for large enough~$N$. In particular, there exist some threshold~$n$ 
for which one can beat the no-go bound via adaptive strategies. As an  example, given $N=10^3$ and $\theta_{\rm max}/\theta_{\rm min}=10$ in Eq.~\eqref{eq:prior} (with $\alpha=1$), adaptive strategies using $n\approx 10$ interacting qubits outperform arbitrary non-adaptive strategies. 

Next, we ask whether the adaptive strategy can reach the Heisenberg-like scaling, $\text{\footnotesize{EMSLE}}^{-1} \propto  mn^2 $. 
To this aim, we study the behaviour of the error with the resources $n$ for a sufficiently large number of repetitions $m$. The results are depicted in Fig.~\ref{fig:error_vs_groupedin}, where we see Eq.~\eqref{eq:ultimate_main} is saturated and therefore the proposed adaptive scheme reaches the ultimate bound on thermometry. 

Finally, we note that although the optimal protocol requires a very idealised Hamiltonian for the probe (a $(2^{n}-1)$-degenerate two-level system),  adaptive protocols already become useful for small $n$. Namely for $n=1,2$, they decrease the error more than $60\%$ and $80\%$, respectively compared to the non-adaptive protocols (see SM for details). For larger $n$, a realistic method to obtain a scaling of the EMSLE beyond the SNL  would be to combine the adaptive method derived here with thermal phase transitions~\cite{huang2009introduction}.

{\it Conclusions and future directions.---}We derived fundamental limitations of the Bayesian approach to equilibrium thermometry, which shows a Heisenberg-like quadratic scaling with the number of probes. We showed non-adaptive strategies cannot saturate this bound and, are limited to shot-noise-like scaling whenever the initial prior is not sharp. We also constructed an adaptive protocol that saturates the ultimate bound, thus highlighting the crucial role of adaptivity in quantum thermometry. This is importantly different to Bayesian phase-estimation protocols~\cite{wiseman2000}, where the Heisenberg limit that applies to most general adaptive protocols~\cite{PhysRevLett.124.030501} can be attained by resorting only to measurements being adaptively varied in between the phase-encoding channel uses~\cite{wiseman_adaptive_2009}. In contrast, in equilibrium thermometry the form of probe states (Gibbs) and measurement (energy-basis) is fixed, and it is the probe Hamiltonian that must be adaptively adjusted for the quadratic scaling to become reachable.

While here we considered the total number of probes $N$ as our resource, future works could include time as an extra resource. This naturally leads to non-equilibrium thermometry, where the probe is measured before reaching thermalisation. While considerable progress in this framework has been obtained within the frequentist approach~\cite{Mehboudi_2019,Seah2020,Kiilerich2018,Cavina2018,PhysRevA.91.012331,Rams_2018}, adaptive protocols could be developed following the Bayesian approach pursued here. 
Lastly, exploiting adaptive schemes for other metrological tasks involving criticality and quantum phase transitions~\cite{Frerot2018}, or restrictions such as limited measurement resolution~\cite{Potts2019fundamentallimits,Hovhannisyan2021,PhysRevResearch.2.033394}, can be subject of future work.


{\it Acknowledgements.---}We gratefully thank  J. Rubio and L. A. Correa for fruitful discussions in an early stage of this work. M.M. and M.P-L. acknowledge  financial support from the Swiss National Science Foundation (NCCR SwissMAP and Ambizione grant PZ00P2-186067). J.K. acknowledges the Foundation for Polish Science within the ``Quantum Optical Technologies'' project carried out within the International Research Agendas programme cofinanced by the European Union under the European Regional Development Fund. MRJ and JBB acknowledge support by the Independent Research Fund Denmark.

\bibliography{mybib.bib,bibPaolo.bib}
\appendix
\onecolumngrid

\section{Derivation of the Van Trees inequality}
\subsubsection{Preliminaries}
We consider a continuous Euclidean one-dimensional parameter space $\Lambda\subseteq \mathbb{R}$.
For a Euclidean space, a suitable function measuring the distance between parameter values is the absolute difference, i.e.
\begin{equation}
    \mathcal{D}(\tilde\lambda,\lambda) = |\tilde\lambda - \lambda| \ \ \text{for} \ \ \tilde\lambda, \lambda \in \Lambda .
\end{equation}
Within the Bayesian approach to parameter estimation we start from a prior probability density $p(\lambda)$ over the parameter space $\Lambda$.
The prior probability is updated as measurement data is acquired.
Given measurement data $\textbf{x}_{m} = \left\lbrace x_{1},...,x_{m}\right\rbrace$, Bayes' theorem allows us to express the update at measurement step $k$ as
\begin{gather} \label{eq:single_shot_Bayes}
    p(\lambda | \textbf{x}_{k}) = \frac{p(x_{k}| \lambda, \textbf{x}_{k-1}) p(\lambda | \textbf{x}_{k-1}) }{ p(x_{k}| \textbf{x}_{k-1})} ,
\end{gather}
where in order to compress the notation we let $p(x_{1}| \lambda, \textbf{x}_{0})=p(x_{1}| \lambda)$, $p(\lambda|\textbf{x}_{0})=p(\lambda)$, and $p(x_{1}| \textbf{x}_{0})=p(x_{1})$.
Here, $p(x_{k}| \lambda, \textbf{x}_{k-1})$ is the likelihood function associated with the implemented measurement. Note that this might be conditional on past measurement outcomes.
We have also defined the marginal density
\begin{equation}
    p(x_{k}| \textbf{x}_{k-1}) \coloneqq \int d\lambda \ p(x_{k}| \lambda, \textbf{x}_{k-1}) p(\lambda | \textbf{x}_{k-1}) .
\end{equation}
For later convenience we introduce the joint probability density $p(\lambda, \textbf{x}_{k}) = p(x_{k}| \lambda, \textbf{x}_{k-1}) p(\lambda , \textbf{x}_{k-1})$, where it is understood that $p(\lambda , \textbf{x}_{0})=p(\lambda)$, and therefore $p(\textbf{x}_{0})=1$ and $p(\textbf{x}_{0}|\lambda)=1$.
Applying equation~\eqref{eq:single_shot_Bayes} iteratively, we can write the posterior distribution resulting from the full measurement trajectory as
\begin{gather}
    p(\lambda | \textbf{x}_{m}) = \frac{p(\textbf{x}_{m}| \lambda) p(\lambda) }{ p(\textbf{x}_{m})} ,
\end{gather}
where we have defined
\begin{gather} \label{eq:likelihood_marginal}
    p(\textbf{x}_{m}| \lambda) = \prod_{k=1}^{m} p(x_{k}| \lambda, \textbf{x}_{k-1}) \\
    p(\textbf{x}_{m}) = \prod_{k=1}^{m} p(x_{k}| \textbf{x}_{k-1}) .
\end{gather}

\subsubsection{Mean-square distance estimation}
An estimation theory is a prescription for specifying a parameter estimate $\bar{\lambda}(\textbf{x}_{m})$,
computed as a function of the data, and for providing a measure of the confidence in the computed estimate.
Here we employ the framework of \textit{mean-square distance} ($\text{\footnotesize{MSD}}$) estimation, in which the confidence in an estimate is gauged by the posterior $\text{\footnotesize{MSD}}$:
\begin{equation} \label{eq:MSD}
\begin{aligned}
    \text{\footnotesize{MSD}}(\textbf{x}_{m}) 
    & \coloneqq \int d\lambda \ p(\lambda | \textbf{x}_{m}) \mathcal{D}(\bar\lambda(\textbf{x}_{m}),\lambda)^{2} , \\
    & = \int d\lambda \ p(\lambda | \textbf{x}_{m}) \ | \bar{\lambda}(\textbf{x}_{m}) - \lambda |^{2} ,
\end{aligned}
\end{equation}
where the second equality follows as we are considering a Euclidean parameter space.
Given a measure of the confidence in an estimate, it is natural to find the maximum confidence estimator.
It can be shown, by minimizing Eq.~\eqref{eq:MSD} with respect to $\bar{\lambda}(\textbf{x}_{m})$,
that the choice of estimator which minimizes the posterior $\text{\footnotesize{MSD}}$ is the posterior mean
\begin{equation}
    \bar{\lambda}(\textbf{x}_{m}) = \int d\lambda \ p(\lambda | \textbf{x}_{m}) \lambda .
\end{equation}
In what follows, we exclusively consider the posterior mean, i.e. the \textit{minimal mean-square distance} ($\text{\footnotesize{MMSD}}$) estimator.
Since the $\text{\footnotesize{MSD}}$ is a stochastic quantity defined for a single measurement trajectory,
it is common to consider the \textit{expected mean-square distance} ($\text{\footnotesize{EMSD}}$):
\begin{equation}
    \text{\footnotesize{EMSD}} := \int d\lambda d\textbf{x}_{m} \ p(\lambda , \textbf{x}_{m}) \ | \bar{\lambda}(\textbf{x}_{m}) - \lambda |^{2} ,
\end{equation}
which is obtained by averaging the $\text{\footnotesize{MSD}}$ over the marginal distribution $p(\textbf{x}_{m})$.
As the name suggest, this quantity gives the $\text{\footnotesize{MSD}}$ which would be obtained on average, if the true parameter value is sampled from the prior probability.

\subsubsection{The Van-Trees inequality}
We now give a derivation of a Bayesian Cram\'{e}r-Rao bound on the $\text{\footnotesize{EMSD}}$, in particular we consider the \textit{Van Trees inequality}~\cite{van2004detection}.
To derive the Bayesian bound we first define the quantity
\begin{equation} \label{eq:H_definition}
\begin{aligned}
    \mathcal{H} \coloneqq \int d\lambda d\textbf{x}_{m} \ & \sqrt{p(\lambda , \textbf{x}_{m})} \ ( \bar\lambda(\textbf{x}_{m}) - \lambda ) \\
    & \times \sqrt{p(\lambda , \textbf{x}_{m})} \  \partial_{\lambda} \log p(\lambda , \textbf{x}_{m})  ,
\end{aligned}
\end{equation}
where differentiability of the posterior distribution is implicitly assumed.
The above quantity is clearly defined to motivate an application of the Cauchy-Schwarz inequality.
From a direct evaluation of the above integral we find
\begin{equation}
\begin{aligned}
    \mathcal{H}
    & = \int d\lambda d\textbf{x}_{m} \ ( \bar\lambda(\textbf{x}_{m}) - \lambda ) \partial_{\lambda} p(\lambda , \textbf{x}_{m}) \\
    & = 1 + \int d\textbf{x}_{m} \left\lbrace \left[ \bar{\lambda}(\textbf{x}_{m}) - \lambda \right] p(\lambda , \textbf{x}_{m}) \right\rbrace_{\lambda \in \mathcal{B}(\Lambda)}  ,
\end{aligned}
\end{equation}
where $\mathcal{B}(\Lambda)$ denotes the boundaries of the parameter space $\Lambda$.
In most cases of interest the boundary term vanish.
Here we take the vanishing of the boundary term as a constraint on the class of models considered, i.e.
\begin{gather}
    p(\lambda , \textbf{x}_{m}) = 0 \ \ \text{for} \ \ \lambda \in \mathcal{B}(\Lambda) \\
    \lambda p(\lambda , \textbf{x}_{m}) = 0 \ \ \text{for} \ \ \lambda \in \mathcal{B}(\Lambda) .
\end{gather}
Given these boundary conditions it follows that $\mathcal{H}=1$.
If we return to the definition of $\mathcal{H}$, i.e. equation~\eqref{eq:H_definition}, and apply the Cauchy-Schwarz inequality, then we obtain the Van Trees inequality:
\begin{equation} \label{eq:gosh_bound}
\begin{aligned}
    \text{\footnotesize{EMSD}}^{-1}
    & \leqslant \int d\lambda d\textbf{x}_{m} \ p(\lambda , \textbf{x}_{m})\left[ \partial_{\lambda} \log p(\lambda , \textbf{x}_{m} ) \right]^{2} \\
    & = Q[p(\lambda)] + \int d\lambda d\textbf{x}_{m} \ p(\lambda,\textbf{x}_{m})  \left[ \partial_{\lambda} \log p(\textbf{x}_{m}|\lambda) \right]^{2} ,
\end{aligned}
\end{equation}
where the second equality follows directly from the decomposition $p(\lambda,\textbf{x}_{m}) = p(\textbf{x}_{m}|\lambda) p(\lambda)$ of the joint probability distribution,
and we have defined the so-called Bayesian information of the prior distribution (which quantifies the prior information about the parameter $\lambda$) as~\cite{van2004detection}:
\begin{equation}
    Q[p(\lambda)] \coloneqq \int d\lambda \ p(\lambda) \left[ \partial_{\lambda} \log p(\lambda) \right]^{2} .
\end{equation}
We can put the Van Trees inequality into the form employed in the main text by decomposing the likelihood function using equation~\eqref{eq:likelihood_marginal},
and then rewriting the expression using Bayes theorem:
\begin{equation} \label{eq:gosh_bound}
\begin{aligned}
    \text{\footnotesize{EMSD}}^{-1}
    & \leqslant Q[p(\lambda)] + \int d\lambda d\textbf{x}_{m} \ p(\lambda,\textbf{x}_{m}) \sum_{k=1}^{m} \left[ \partial_{\lambda} \log p(x_{k} |\lambda, \textbf{x}_{k-1}) \right]^{2} \\
    & = Q[p(\lambda)] + \sum_{k=1}^{m} \int d\lambda d\textbf{x}_{k-1} \ p(\lambda,\textbf{x}_{k-1})
    \int dx_{k} p(x_{k} |\lambda, \textbf{x}_{k-1}) \left[ \partial_{\lambda} \log p(x_{k} |\lambda, \textbf{x}_{k-1}) \right]^{2} \\
    & = Q[p(\lambda)] + \sum_{k=1}^{m} \int d\textbf{x}_{k-1} p(\textbf{x}_{k-1}) \int d\lambda \ p(\lambda |\textbf{x}_{k-1}) \ h_{k}(\lambda) \\ 
\end{aligned}
\end{equation}
where
\begin{equation}\label{eq:Fisher_info}
    h_{k}(\lambda) \coloneqq \int dx_{k} p(x_{k} |\lambda, \textbf{x}_{k-1}) \left[ \partial_{\lambda} \log p(x_{k} |\lambda, \textbf{x}_{k-1}) \right]^{2} .
\end{equation}
is just the Fisher Information of the distribution $p(x_{k} |\lambda, \textbf{x}_{k-1})$ evaluated with respect to the parameter $\lambda$~\cite{van2004detection}. 
Note that the Fisher information $h_{k}(\lambda)$ is generally conditioned on the past measurement trajectory $\textbf{x}_{k-1}$---a fact that we suppress in the notation for simplicity.

\section{Application to equilibrium thermometry}
\subsubsection{Preliminaries}
We now turn our attention to equilibrium probe thermometry.
Let $\theta\in\Theta$ denote the sample temperature, where $\Theta$ is the space of temperatures.
We consider a measurement consisting of first thermalizing the $n$-qudit probe system, described by a Hamiltonian operator $H_{n}^{(k)}$,
and then performing a projective energy measurement of the probe.
The probe at measurement step $k$ is found in the thermal Gibbs state
\begin{align}
    \omega(\theta;H_{n}^{(k)}) = \frac{e^{- H_{n}^{(k)}/k_{B}\theta}}{\mathcal{Z}_{n}^{(k)}}
\end{align}
with $\mathcal{Z}_{n}^{(k)}={\rm Tr} (e^{- H_{n}^{(k)}/k_{B}\theta})$, and $k_{B}$ is Boltzmann's constant.
For convenience, the ground-state energy is set to zero.
The definitions of the average probe energy and the probe heat capacity are:
\begin{gather}
   E(\theta; H_{n}^{(k)}) \coloneqq {\rm Tr}(H_{n}^{(k)} \omega(\theta;H_{n}^{(k)}) ) , \\
    C(\theta; H_{n}^{(k)}) \coloneqq \frac{{\rm d}  E(\theta; H_{n}^{(k)}) }{{\rm d}  \theta}.
\end{gather}

\subsubsection{Mean-square logarithmic error}
The $\text{\footnotesize{MSD}}$ estimation theory developed in the preceding sections is defined with respect to a Euclidean parameter space $\Lambda$.
In the case of equilibrium probe thermometry, the space of temperatures is not a Euclidean parameter space.
However, in the specific case of equilibrium probe thermometry of a thermalizing channel,
the space of temperatures can be mapped into a Euclidean space by taking the logarithm~\cite{rubio2020global,Mathias_accompany}
\begin{equation}
    \lambda(\theta) = \log(\theta) .
\end{equation}
The $\text{\footnotesize{EMSD}}$ then takes the form of an \textit{expected mean-square logarithmic error} ($\text{\footnotesize{EMSLE}}$) studied in the main text.

\subsubsection{Van Trees inequality in thermometry}
For the sake of generality we will stick to an arbitrary parameterization $\lambda(\theta)$,
i.e. a one-to-one map $\lambda:\Theta\rightarrow \Lambda$, which is assumed to be differentiable.
The Fisher information transforms under a change of parameterization as
\begin{equation}
    h_{k}(\lambda) = (d\theta/d\lambda)^{2} h_{k}(\theta) .
\end{equation}
Here, the data is obtained via projective energy measurements of the probe system.
In fact, this is the optimal measurement maximising the Fisher information, which thus constitutes then the so-called \textit{quantum Fisher information} being directly related to 
the heat capacity of the probe, i.e.~\cite{correa2015individual}:
\begin{gather}
    C(\theta;H_{n}^{(k)}) = \theta^{2} h_{k}(\theta) ,
\end{gather}
which is a functional of the probe Hamiltonian.
In terms of the probe heat capacity the posterior averaged Fisher information introduced in the preceding section takes the form
\begin{equation}
    \text{\footnotesize{EMSD}}^{-1} \leqslant Q[p(\theta)] + \sum_{k=1}^{m} \int d\textbf{x}_{k-1} p(\textbf{x}_{k-1}) \int d\theta \ p(\theta |\textbf{x}_{k-1}) \ 
    \left[ \frac{1}{\theta}\frac{d\theta}{d\lambda} \right]^{2} C(\theta;H_{n}^{(k)}) .
\end{equation}
where we have made use of the parameterization invariance of the probability density, i.e. $d\lambda p(\lambda | \textbf{x}) = d\theta p(\theta | \textbf{x})$,
which is a requirement on a well-defined probability density.
For convenience, we define
\begin{equation}
    J_{\lambda}(\theta) \coloneqq \left[ \frac{1}{\theta}\frac{d\theta}{d\lambda} \right],
\end{equation}
and note that in the specific case of $\lambda(\theta) = \log(\theta)$ it follows that $J_{\lambda}(\theta)=1$, in which case we recover the form of the Van Trees inequality given in the main text.
Lastly, we note that when working with the logarithmic parameterization, the Bayesian information of the prior takes the form
\begin{equation}
    Q[p(\theta)] = \int d\theta\, p(\theta) \left[ 1 + \partial_{\theta} \log p(\theta) \right]^{2}
\end{equation}

We are interested in the optimal probe design, and formally define the optimization problem
\begin{equation} \label{eq:gamma_opt}
\begin{aligned}
    \Gamma(\textbf{x}_{k-1}) \coloneqq   \max_{H_{n}^{(k)}} \int d\theta \ p(\theta | \textbf{x}_{k-1})J_{\lambda}(\theta)^{2} C(\theta;H_{n}^{(k)})
\end{aligned}
\end{equation}
Note that $\Gamma$ is in general a functional of the past measurement trajectory, i.e., the optimal probe structure depends on the prior knowledge of the parameter to be estimated.
In the following sections we derive model-independent upper bounds on $\Gamma$.

\subsubsection{Model-independent super-extensive upper bound on $\Gamma$}
In this section we derive a super-extensive bound on $\Gamma(\textbf{x})$.
Starting with Eq. \eqref{eq:gamma_opt}, we note that since the integrand is positive we can provide an upper bound by moving from a global maximization to a local maximization, i.e.
\begin{equation}
    \Gamma(\textbf{x}) \leqslant \int d\theta \ p(\theta | \textbf{x}) J^{2}_{\lambda}(\theta) \ \max_{H} \ C(\theta;H) .
\end{equation}
The problem of maximizing the heat capacity, over all possible probe Hamiltonians at a given temperature, has been solved by Correa et al.~\cite{correa2015individual}.
The solution can be formulated as the temperature-independent tight upper bound
\begin{equation}
    C(\theta;H) \leqslant \left[ \frac{\xi_{D}}{2} \right]^{2} - 1 ,
\end{equation}
where $\xi_{D}$ is the solution to the transcendental equation
\begin{equation}
    e^{\xi_{D}} = \left( D-1 \right) \frac{\xi_{D}+2}{\xi_{D}-2} .
\end{equation}
This equation does not have a closed form solution. However, a general feature of the solution is that $\xi_{D} > \log(D-1)$, and that $\xi_{D}$ approach $\log(D-1)$ from above as $D$ becomes large.
From this it follows that $\Gamma(\textbf{x})$ satisfies the super-extensive upper bound
\begin{equation}
\begin{aligned}
    \Gamma(\textbf{x}) 
    & \leqslant \left(\xi_{D}/2 - 1\right) \left(\xi_{D}/2 + 1\right) \int d\theta \ p(\theta | \textbf{x}) J^{2}_{\lambda}(\theta) ,
\end{aligned}
\end{equation}
which grows super-extensively in $\log(D)$.
If we average $\Gamma(\textbf{x})$ over the past measurement trajectory we find
\begin{equation} \label{eq:super_extensive_bound}
    \int d\textbf{x} p(\textbf{x})\Gamma(\textbf{x}) \leqslant \left(\xi_{D}/2 - 1\right) \left(\xi_{D}/2 + 1\right) \left\langle J^{2}_{\lambda}\right\rangle_{\text{prior}}
    \equiv C_{D} \left\langle J^{2}_{\lambda}\right\rangle_{\text{prior}},
\end{equation}
where we have defined
\begin{equation}
    \left\langle J^{2}_{\lambda}\right\rangle_{\text{prior}} = \int d\theta \ p(\theta) J^{2}_{\lambda}(\theta) .
\end{equation}
This bound is expected to be approximately tight in the limit where the prior is local with respect to the width of the heat capacity.
As we will see in the next section, designing a probe with a critical heat capacity at a certain temperature, i.e. one attaining the maximal heat capacity,
will result in the width of the heat capacity decreasing as $1/\log(D)$.
We thus see that saturating the super-extensive bound requires a prior probability distribution confined to a domain 
$\theta \in \left[ \theta_{c}-\Delta/2 , \theta_{c}+\Delta/2 \right]$ where $\theta_{c}$ is the critical temperature and $\Delta = 1/\log(D)$.
As $D$ increase this corresponds to an increasing amount of prior information.

\subsubsection{Tight upper bound on the thermal energy density}
In this section we want to derive an upper bound on the thermal energy at a given temperature for any probe structure, subject to the dimensionality constraint $\dim{H}=D$ on the considered probes.
We will find that the thermal energy density is upper bounded by the temperature.
Define the maximum thermal energy for any probe structure as
\begin{equation}
    E_{\max}(\theta) \coloneqq \max_{H} \ E(\theta; H) ,
\end{equation}
\begin{equation}
    E(\theta; H) \coloneqq \Tr{ H \omega(\theta;H) } ,
\end{equation}
where $\omega(\theta;H)$ is a thermal state at temperature $\theta$.
We denote the energy eigenvalues of the probe Hamiltonian by $\left\lbrace \varepsilon_{l} \right\rbrace$, and for convenience set the ground-state energy to zero.
If we take the derivative of the thermal energy, and equate to zero we obtain the condition
\begin{equation}
    \varepsilon_{l} = \theta + E(\theta;H)\eqqcolon \varepsilon ,
\end{equation}
which implies a $D-1$ degeneracy in the first excited state.
Evaluating the above condition for this probe structure leads to a transcendental equation for $\varepsilon/\theta$ which can be solved.
The result is the temperature-dependent upper bound
\begin{gather} \label{eq:energy_density_bound}
    E(\theta;H) \leqslant \theta \mathcal{W}_{D} \\
    \mathcal{W}_{D} \coloneqq  W\left(\frac{D-1}{e} \right) ,
\end{gather}
where $W$ denotes the product logarithm, also called the \textit{Lambert W} function.
In the limit of large $D$ the behaviour of the product logarithm is such that $\mathcal{W}_{D}$ tends asymptotically to $\log(D)$ from below.
We stress that the above bound on the thermal energy can be saturated by an effective two level probe with a $D-1$ degenerate excited state, and a temperature-dependent energy gap.

\subsubsection{Extensive bound for the non-adaptive scenario}
W start with the second term in Eq~(4) of the main text. Since the Hamiltonian remains constant throughout the protocol, i.e., $H_n^{(k)}=H_n$ $\forall k$,  this term can be rewritten as
\begin{align}
    {\bar \Gamma} 
    & \coloneqq \sum_{k=1}^{m} \iint\! d\theta d\textbf{x}_{k-1}\,p(\theta) \,p(\textbf{x}_{k-1}|\theta)\,C(\theta;H_n) \nonumber\\
    & = m \int\!d\theta \, p(\theta) \,  C(\theta;H_n).
    \label{eq:Lambda}
\end{align}
Integrating by parts---recall that $C(\theta;H_n)=\partial_{\theta} E(\theta;H_n)$---and maximising over  $H_n$ gives
\begin{equation}
\label{eq:Lambda_UB}
    {\bar \Gamma} \leqslant m\max_{H_n} \int d \theta \ [-\partial_{\theta}p(\theta)] E(\theta;H_n),
\end{equation}
where we assumed that $p(\theta) E(\theta;H_n)$ is smooth and vanishes at the boundaries. By defining ${\cal R}$ as the temperature domain where $\partial_{\theta}p(\theta)\leqslant 0$ we have
\begin{align}
\label{eq:derivingnogo}
    {\bar \Gamma} & \leqslant m\max_{H_n}\int_{\cal R} d \theta \ [-\partial_{\theta}p(\theta)] E(\theta;H_n)\nonumber\\
    & \leqslant m \int_{\cal R} d \theta \ [-\partial_{\theta}p(\theta)] \max_{H_n}E(\theta;H_n)
\end{align}
To make further progress, we use the upper bound on the energy of an $n$-body system at thermal equilibrium (with total dimension $D=d^n$) that is given by Eq.~\eqref{eq:energy_density_bound}:
\begin{align}
    \max_{H_n}E(\theta;H_n) & \leqslant \theta \hspace{0.5mm} {\cal W}_{D} \leqslant \theta \hspace{0.5mm} n\log d
    \label{eq:upperBoundE}
\end{align}
where 
the second equality is saturated as $n\gg 1$. 
Plugging these results back into Eq. (4) of the main text we obtain  
a  no-go theorem for non-adaptive strategies [Result (ii)]
\begin{align}\label{eq:non_adapt_nogo}
 \hspace{-1mm} \text{\footnotesize{EMSLE}}^{-1}
        \overset{\mathrm{non-adaptive}}{\leqslant} Q[p(\theta)] + f[p(\theta)] mn\log\!d, 
\end{align}
where $f[p(\theta)] = \int_{\cal R} d \theta \ [-\partial_{\theta}p(\theta)] \theta$ is a functional of the prior.
Crucially, the bound \eqref{eq:non_adapt_nogo} implies that, even with arbitrary control over the $n$-body Hamiltonian, one cannot go above a linear scaling in $n$ with non-adaptive strategies (compare with the general bound given by Eq.~(7) of the main text).
%
%
%

Our alternative bound follows the exact same procedure, except we first recall that the thermal energy can be expressed as $\theta^{2} \partial_{\theta}\Psi(\theta;H) $, where the  Massieu potential
reads $\Psi(\theta;H) \coloneqq \log \mathcal{Z}(\theta;H)$ with $\mathcal{Z}(\theta;H)$ being the partition function of the probe.
Starting from Eq.~\eqref{eq:Lambda} and by performing twice integration by parts we get
\begin{align}
    {\bar \Gamma} 
    &= m \int\!d\theta \, p(\theta) \,  C(\theta;H_n) = - m\int d\theta \ \left[ \partial_{\theta}   p(\theta)) \right] E(\theta;H)\nonumber\\
    &= \ m\int d\theta \ \left[ \partial_{\theta} \left(\theta^{2} \partial_{\theta} p(\theta)\right) \right] \Psi(\theta;H),
\end{align}
where again we take the vanishing and differentiablity of the boundary terms in both integrations---that is $p(\theta) E(\theta;H_n)$ and $\theta^{2} \partial_{\theta} p(\theta) \Psi(\theta;H)$---as a restriction on the choice of parameterization.
We can derive an upper bound on the optimal solution by noting that $\Psi(\theta;H)\geqslant 0 $---recall that the ground state energy is set to zero---and by introducing
$\bar{\mathcal{R}}=\left\lbrace \theta \mid  \partial_{\theta} \left(\theta^{2} \partial_{\theta} p(\theta)\right)  \geqslant 0 \right\rbrace$. Then
\begin{equation}
\begin{aligned}
    {\bar \Gamma}
    & \leqslant m\max_{H} \ \int_{\bar{\mathcal{R}}} d\theta \ \left[ \partial_{\theta} \left(\theta^{2} \partial_{\theta} p(\theta)\right) \right] \Psi(\theta;H) .
\end{aligned}
\end{equation}
As the integrand is now positive we can maximize the Massieu potential locally.
Since the logarithm is monotonically increasing in its argument, this corresponds to substituting the largest value of the partition function, i.e. the Hilbert space dimension.
The bound then takes the form
\begin{equation}
\begin{aligned}
    {\bar \Gamma}
    & \leqslant m \log(D) \int_{\bar{\mathcal{R}}} d\theta \ \left[ \partial_{\theta} \left(\theta^{2} \partial_{\theta} p(\theta)\right) \right]\\
    & \leqslant m\log(D) \left\lbrace \theta^{2} \partial_{\theta} p(\theta) \right\rbrace_{\bar{\mathcal{R}}} \\    
    & \eqqcolon m\log(D) g[p(\theta)] ,
\end{aligned}
\end{equation} 
where $g[p(\theta)]$ is a functional of the prior distribution but independent of the probe.
This gives two complementary bounds on ${\bar \Gamma}$, i.e. one expressed in terms of $f[p(\theta)]$ as presented in the main text, and one in terms of $g[p(\theta)]$.
Which of these two is tighter depends on the specific prior.
\section{The ${\rm EMSLE}$ for small number of interacting qubits: adaptive vs non-adaptive}
%
\begin{figure}[H]
    \centering
    \includegraphics[width=.5\linewidth]{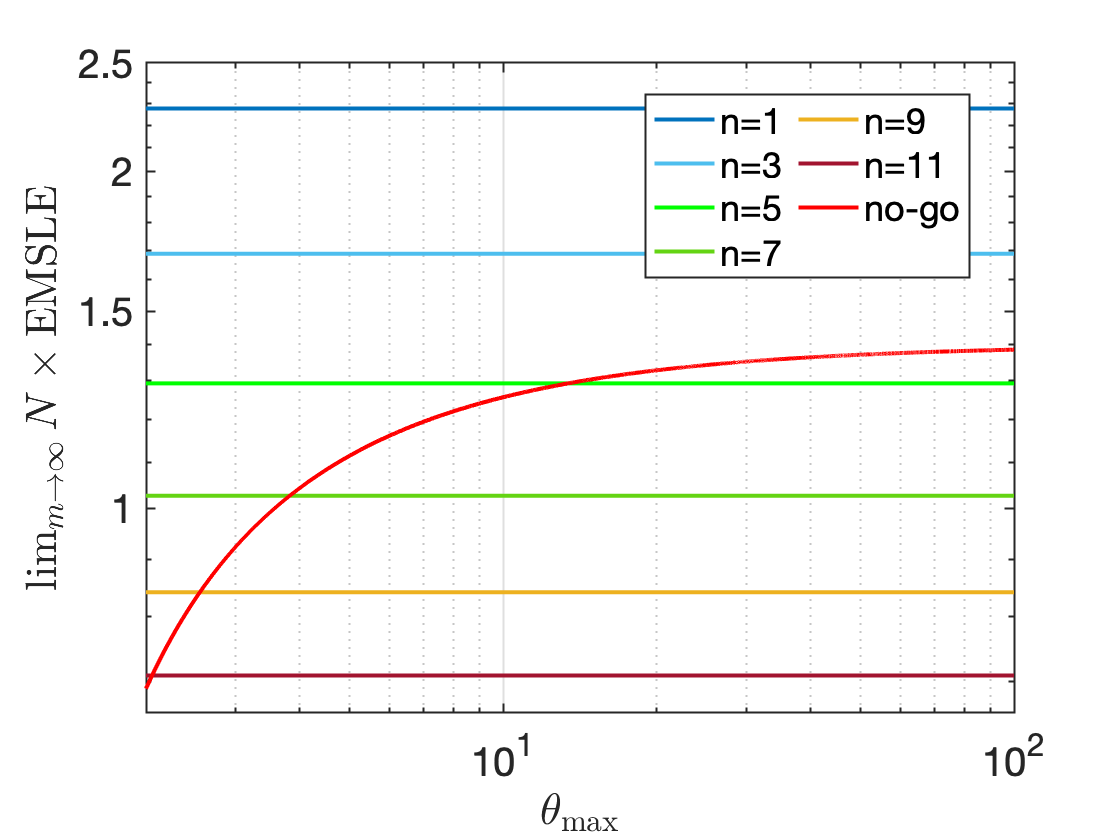}
    \caption{The asymptotic value of normalised error, $\lim_{m\to\infty} N \times {\rm EMSLE}$ (here $m=2\times 10^{5}$) vs the maximum temperature in the prior $\theta_{\max}$, for the non-adaptive no-go theorem given by Eq.~(12) of the main text (red curve). The horizontal lines show the same quantity in the adaptive scenario when using interacting qubits. One can see that for $\theta_{\max} \geqslant 20$ and by setting $n\geq 5$ the adaptive strategy overperforms any non-adaptive strategy.
    Here we set $\theta_{\min}=1$ and $\alpha=-20$.}
    \label{fig:adaptive_vs_nogo}
\end{figure}
In the main text we demonstrated that by choosing $n>10$ the adaptive strategy can reach a precision that any non-adaptive counterpart cannot reach (Fig.~2 of the main text, right panel). 
The exact value of interacting qubits $n$ for which the adaptive strategy beats the non-adaptive no-go bound depends on the prior. For instance, in Fig.~\ref{fig:adaptive_vs_nogo} we see that for some priors, adaptive strategies with $n=5$ can beat the no-go theorem. We also emphasize  that the  no-go bound is not necessarily tight, in practice  non-adaptive strategies might be far from them.

Nonetheless, one might still wonder about the experimental preparation of effectively two level probes with maximally degenerate excited state. In an upcoming paper, some of us show that similar energy structures can be prepared with spin Hamiltonians that contain only two-body interactions \cite{Michael&Marti&others}. Yet still, our adaptive scheme is advantageous even in a single qubit or two qubits scenario (i.e., $n\in \{1,2\}$), with two effective energy levels and a tunable gap.
%
\begin{figure}[H]
    \centering
    \includegraphics[width=.43\linewidth]{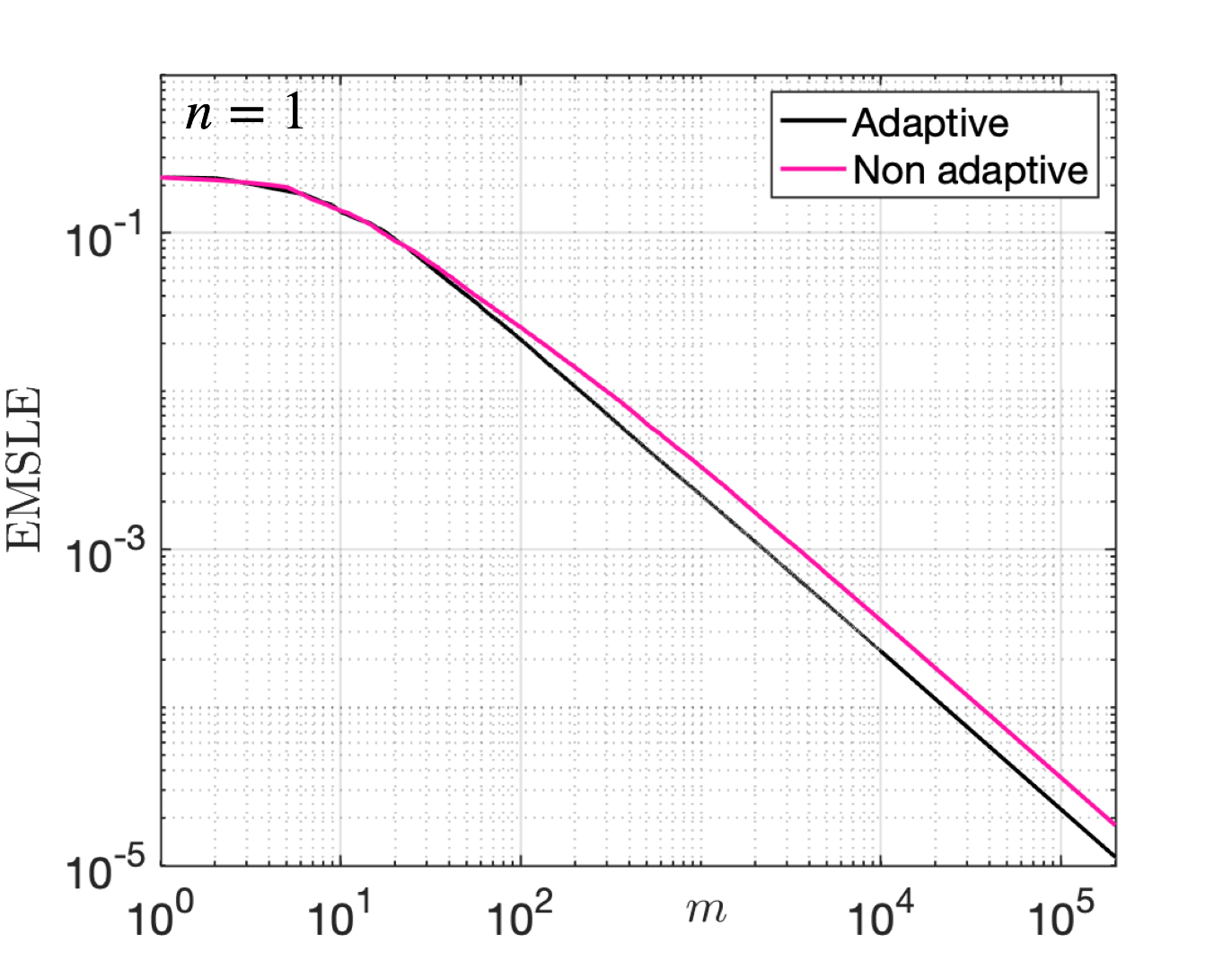}
        \includegraphics[width=.43\linewidth]{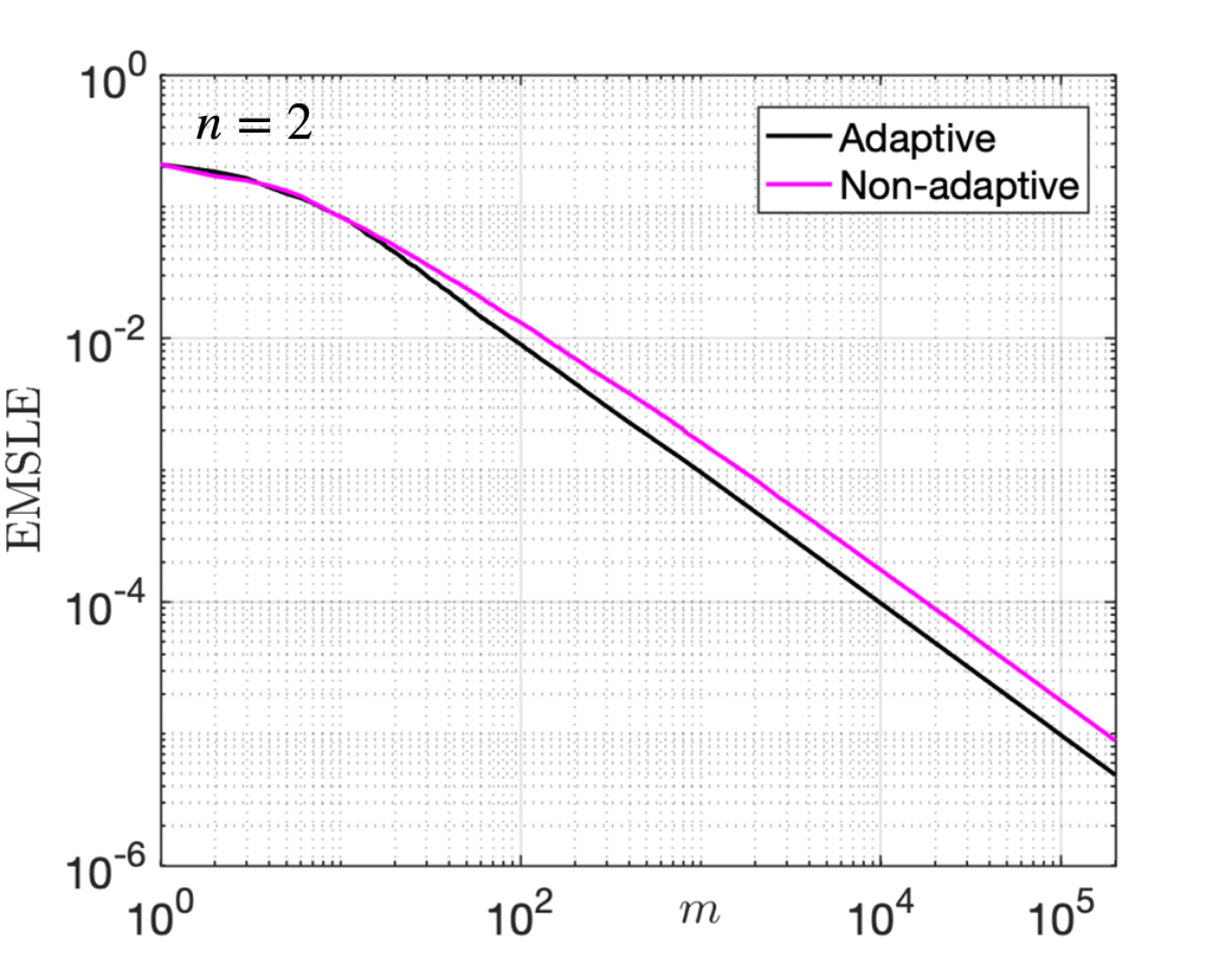}
    \caption{The illustration of superiority of adaptive strategies vs non-adaptive ones. 
    Clearly for sufficently large repetitions the adaptive scheme over performs the non-adaptive one. In particular, 
    to reach the same target error ${\rm EMSLE}={\cal O}(2\times 10^{-5})$ one needs about $60\%$ less repetitions in the adaptive scenario compared to the non-adaptive scheme for $n=1$ as depicted on the left panel (similarly, one needs about $80\%$ less measurements for $n=2$, as depicted on the right panel). Here the prior is the same as Eq.~(13) of the main text with $\theta_{\min}=1$, $\theta_{\max}=10$, and $\alpha=-20$.}
    \label{fig:adaptive_vs_non}
\end{figure}
As illustrated in Fig.~\ref{fig:adaptive_vs_non} one sees that to reach the same target error (${\rm EMSLE}={\cal O}(2\times 10^{-5})$), the adaptive scheme requires roughly $60\%$ less measurement runs compared to the non-adaptive strategy for $n=1$, while for $n=2$ the adaptive strategy requires roughly $80\%$ less measurement runs. 
\section{The non-asymptotic {\rm EMSLE}}
\begin{figure}
    \centering
    \includegraphics[width=.5\linewidth]{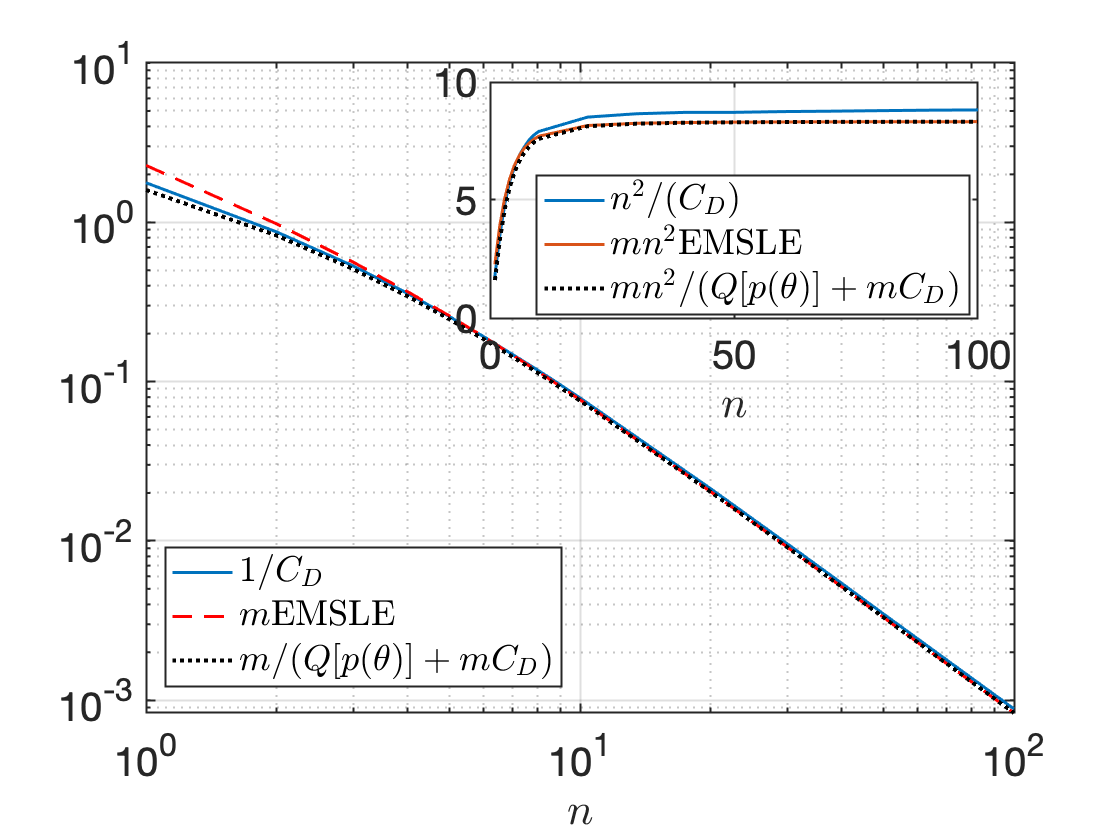}
    \caption{The non-asymptotic behaviour of the ${\rm EMSLE}$ (here $m=100$). Inspecting the inset above, one can see that, unlike the high $m$ regime here we have $m{\rm EMSLE}\neq 1/C_D$. This is so because the term $Q[p(\theta)]$  still plays a prominent role. In fact, if we incorporate this term, the inequality (7) of the main text will reduce to equality even for such a small number of repetitions.}\label{fig:Heisenberg_lowm}
\end{figure}

In the main text we demonstrated how our proposed adaptive scheme can saturate the ultimate bound Eq.~(7) of the main text, as depicted in Fig.~3 of the main text. The saturability of the bound is gauranteed by choosing high enough number of repetitions ($m=2000$ in the main text). In case we were to perform less measurements, the bound is not generally saturated. Moreover, the first term in the r.h.s. of Eq.~(7), i.e., $Q[p(\theta)]$ will also play a role. This is depicted in Fig.~\ref{fig:Heisenberg_lowm}.
\end{document}